\documentclass[aps,pre,superscriptaddress,showpacs,twocolumn]{revtex4}

\usepackage{amsmath,amssymb}
\usepackage{csquotes}
\usepackage{enumitem}
\usepackage{tikz}
\usetikzlibrary{shadings,patterns}
\usepackage[capitalize]{cleveref}

\usepackage{mathtools}
\DeclarePairedDelimiter\abs{\lvert}{\rvert}

\renewcommand{\vec}[1]{\mathbf{#1}}			
\newcommand{\fourier}[1]{\hat{#1}}				
\newcommand{\vort}{\boldsymbol{\omega}}		
\newcommand{\partfun}{\ensuremath{\mathcal{Z}}}	
\newcommand{\torus}{\mathbb{T}}
\newcommand{\toruscal}{\mathcal{T}}
\newcommand{\pdual}[1]{\hat{#1}}				

\begin{document}

\title{Restricted Partition Functions and Inverse Energy Cascades\\
 in Parity Symmetry Breaking flows}

\author{Corentin Herbert}
\email{cherbert@ucar.edu}
\affiliation{National Center for Atmospheric Research, P.O. Box 3000, Boulder, CO, 80307, USA}

\begin{abstract}
When the symmetries of homogenous isotropic turbulent flows are broken, different sets of modes with different physical roles emerge. In particular, choosing a forcing which puts more weight on one or the other of these sets may result in different statistics for the energy transfers. We use the general method of computing a partition function restricted to a portion of phase space to study analytically these different statistics. We illustrate this method in the case of parity symmetry breaking, measured by helicity. It is shown that when helicity is sign definite at all scales, an inverse cascade is expected for the energy.
 When sign-definiteness is lost, even for a small set of modes, this cascade disappears and there is a sharp phase transition to the standard helical equipartition spectra.
\end{abstract}


\maketitle

\section{Introduction}

Turbulent flows are commonly described in terms of their ability to transfer energy from one scale of motion to the other. The dimension of the domain plays a fundamental role in this process and two major phenomenologies are known: in 3D isotropic-homogeneous flows, the energy is preferentially transferred from the large to the small scales in a process referred to as the \emph{Kolmogorov-Richardson cascade} \cite{Kolmogorov1941a,FrischBook}. By contrast, 2D flows transfer energy from the small scales to the large scales, as the name of \emph{inverse cascade} indicates \cite{Kraichnan1967}. It appears that these two standard frameworks are in fact very particular cases, and a wider family of behaviors can be obtained by breaking one of the numerous symmetries of such isotropic-homogeneous flows. In realistic flows, like for instance geophysical flows, physical effects like rotation, stratification or simply geometrical confinement break isotropy by imposing a preferred direction. This may result equally well in the emergence of an inverse or direct cascade \cite{Metais1996,LSmith1996,LSmith2002,Celani2010,Aluie2011,Marino2013b}. In fact, the isotropic vocabulary may no longer be adequate, and it may become necessary to describe the energy transfers in terms of the components of the wave vector parallel --- $k_\parallel$ --- or perpendicular --- $k_\perp$ --- to the preferred direction, rather than just in terms of the modulus $k$ of the wave vector.
 Rigorous analysis is made even more complicated by the introduction of a new time scale, in addition to the eddy turnover time, corresponding to the propagation of waves \cite{NazarenkoBook}. The symmetry group for isotropic flows is $O(3)=\mathbb{Z}_2 \times SO(3)$; rotation and stratification break the continuous subgroup $SO(3)$ into an $SO(2)$ symmetry, and lead to
the appearance of waves. On the contrary breaking the discrete subgroup $\mathbb{Z}_2$, or in other words the parity symmetry: $P: \vec{x} \mapsto -\vec{x}$, does not lead to such complications. The extent to which the latter symmetry is broken is measured by an invariant quantity called helicity~\cite{Moreau1961} (it also measures the correlation between velocity and vorticity, or the topology of the vortex lines \cite{Moffatt1969,Moffatt1992a,Moffatt2001}); a parity-invariant flow has vanishing helicity.
In this paper, we will consider 3D turbulent flows with helical constraints, in such a way that the degree with which the parity symmetry is broken, and at which scales it is broken, can be accurately controlled. We show analytically that these constraints can lead to an inverse cascade, without imposing any additional effect like rotation or stratification. A particularly interesting case is that of maximal symmetry breaking, for which velocity and vorticity are aligned (\emph{Beltrami flows}). In this context, the existence of an inverse cascade was shown experimentally~\cite{EricHerbert2012} and numerically~\cite{Biferale2012}, and justified phenomenologically on the basis of an analogy with 2D flows~\cite{Kraichnan1973,Biferale2012}. Here, we provide additional analytical insight through an equilibrium statistical mechanics approach.

To do so, we make use of a refinement of the statistical mechanics approach initiated by Lee \cite{Lee1952} and Kraichnan \cite{Kraichnan1967,Kraichnan1973}. The standard approach consists in building a probability distribution on phase space based only on the invariants of the system. In the canonical framework, the statistics of the system are encoded in the partition function, which is an integral over phase space. In this integral, the contributions from the microstates concentrating around the equilibrium macrostate dominate. 
Thus, in order to study the statistical role of states which are only metastable, one should restrict the integral to a portion of phase space chosen in an appropriate manner, to get rid of the dominant contributions, as suggested by Penrose and Lebowitz~\cite{Penrose1979}. Up to now, this technique has been used mostly in the context of \enquote{toy} models of statistical physics, like the Ising model of ferromagnetism~\cite{Capocaccia1974}, the van der Waals-Maxwell model of the liquid-vapor transition~\cite{Penrose1971} and variants like the Widom and Rowlinson model~\cite{Cassandro1977}. In this paper, we show that this technique is also relevant to investigate the different behaviors embedded in such a complex system as a turbulent flow. Indeed, in many cases, nonlinear interactions transferring energy toward the small and large scales both exist and are almost in balance. The existence of an energy cascade and its direction results from a slight imbalance which may be traced back to the constraints imposed by the conservation laws.
As soon as one breaks slightly the symmetries, the phase space breaks in different sets of modes with a different physical nature; \emph{universality} may even be lost and several phenomenologies may be observed depending on the repartition of the forcing in phase space (e.g. isotropic or anisotropic, in the 2D or the 3D modes, etc) \cite{Sen2012}. Restricted partition functions provide a precise tool to predict theoretically these coexisting possibilities. As an example, we show here that the restricted partition functions obtained by summing over a portion of phase space defined by helical constraints related to parity symmetry breaking may lead to the existence of an inverse cascade for the energy. Since, for such helically constrained flows, the helicity plays an analogous role to enstrophy in 2D flows, we also discuss the differences between the equilibrium properties of these two systems, especially the equilibrium spectrum at large $k$.

In section \ref{3dstatmechsec}, we introduce the notations and recover the classical absolute equilibrium spectra for 3D helical turbulence~\cite{Kraichnan1973} by computing the partition function rather than through equipartition theorems. In section \ref{symbreakflowssec}, we restrict the computation of the partition function to submanifolds defined by helical constraints. Section \ref{maxsymbreaksec} is devoted to Beltrami flows which maximally break the parity symmetry; we describe in details the different regimes which arise at statistical equilibrium and discuss the analogy with 2D flows. In section \ref{infsymbreaksec}, we investigate the effect of weaker helical constraints and show that the condition for the inverse cascade regime to persist is that the relative helicity should be positive definite at all scales.

\section{Partition Function and Absolute Equilibrium for 3D Turbulence}\label{3dstatmechsec}

Ideal (unforced and inviscid) incompressible flows on a three dimensional cubic domain with periodic boundary conditions (i.e. on the torus $\toruscal=\torus^3$) are governed by the Euler equations for the velocity field $\vec{u} \in L^2(\toruscal)$:
\begin{equation}
\partial_t \vec{u} + \vec{u} \cdot \nabla \vec{u} = - \nabla P, \quad \nabla \cdot \vec{u} = 0.
\end{equation}
It is customary to introduce the Fourier decomposition for the velocity field 
\begin{align}
\vec{u}(\vec{x})&=\sum_{\vec{k} \in \pdual{\toruscal}} \vec{\fourier{u}}(\vec{k}) e^{i\vec{k}\cdot\vec{x}}
\intertext{and the vorticity field (defined by $\vort = \nabla \times \vec{u}$)}
\vort(\vec{x}) &= \sum_{\vec{k}\in \pdual{\toruscal}} \fourier{\vort}(\vec{k}) e^{i\vec{k}\cdot\vec{x}}, 
\end{align}
with $\fourier{\vort}(\vec{k})=i \vec{k} \times \vec{\fourier{u}}(\vec{k})$, and $\pdual{\toruscal}=2\pi/L\mathbb{Z}^3$ the Pontryagin dual of the torus $\torus^3$. The condition for the velocity field to be real is that $\vec{\fourier{u}}(-\vec{k})=\vec{\fourier{u}}(\vec{k})^*$.
The incompressibility condition reads $\vec{\fourier{u}}(\vec{k}) \cdot \vec{k}=0$. Hence, the \emph{phase space} of the system is in fact the submanifold 
\begin{align}
\Lambda&=\{ \vec{\fourier{u}} \in L^2(\pdual{\toruscal}), \forall \vec{k} \in \pdual{\toruscal},  \vec{\fourier{u}}(-\vec{k})=\vec{\fourier{u}}(\vec{k})^* \text{ and } \vec{\fourier{u}}(\vec{k}) \cdot \vec{k}=0\}
\end{align}
of $L^2(\pdual{\toruscal}) \simeq L^2(\toruscal)$ --- it is in fact a vector space.
 
 The dynamics in Fourier space reads
 \begin{align}
 \partial_t \fourier{u}_\alpha(\vec{k}) &= - \frac i 2 P_{\alpha\beta\gamma}(\vec{k}) \sum_{\vec{p} \in \pdual{\toruscal}} \fourier{u}^\beta(\vec{p})\fourier{u}^\gamma(\vec{k}-\vec{p}),
 \end{align}
 where $P_{\alpha\beta\gamma}(\vec{k})=k_{\beta}\mathcal{P}_{\alpha\gamma}(\vec{k})+k_{\gamma}\mathcal{P}_{\alpha\beta}(\vec{k}), \mathcal{P}_{\alpha\beta}(\vec{k})=\delta_{\alpha\beta}-k_{\alpha}k_{\beta}/k^2$  is the projection operator which ensures that the flow remains incompressible, i.e. that $\Lambda$ is an invariant manifold for the dynamics, as easily checked.
 
 A useful description of phase space which automatically enforces the incompressibility constraint is the Craya-Herring \cite{Craya1958,Herring1974}
decomposition: for any wavenumber $\vec{k}$, we introduce the eigenvectors of the rotational operator $\vec{h}_\pm(\vec{k})$ such that $i\vec{k} \times \vec{h}_\pm(\vec{k}) = \pm k \vec{h}_\pm (\vec{k})$ \cite{Waleffe1992}. 
In this new basis, we may write 
\begin{align}
\vec{\fourier{u}}(\vec{k}) &= u_+(\vec{k}) \vec{h}_+(\vec{k}) + u_-(\vec{k}) \vec{h}_-(\vec{k}),\\
\fourier{\vort}(\vec{k}) &= k \lbrack u_+(\vec{k}) \vec{h}_+(\vec{k}) - u_-(\vec{k}) \vec{h}_-(\vec{k})\rbrack,
\end{align}
where $u_+(\vec{k})$ and $u_-(\vec{k})$ are arbitrary complex coefficients satisfying only the reality condition for the velocity field: $u_\pm(-\vec{k})=u_\pm(\vec{k})^*$ (note that $\vec{h}_\pm(-\vec{k})=\vec{h}_\pm(\vec{k})^*$). 
This decomposition can be interpreted physically as splitting each plane wave in the Fourier decomposition as the sum of two circularly polarized helical waves, with opposite polarizations.
Now the phase space is simply described by 
\begin{align}
\Lambda &= \{ (u_+,u_-) \in L^2(\pdual{\toruscal})^2,  \forall \vec{k} \in \pdual{\toruscal},  u_\pm(-\vec{k})=u_\pm(\vec{k})^*\}.
\end{align}
In this description, the dynamics reads~\cite{Waleffe1992}:
\begin{equation}
\begin{split}
\partial_t u_s(\vec{k}) = & - \frac 1 4  \sum_{\vec{p}, \vec{q} \in \pdual{\toruscal} } \sum_{s_p,s_q \in \{+,-\}} (s_p p - s_q q) \times \\
& \lbrack (h_{s_p}(\vec{p})^* \times h_{s_q}(\vec{q})^*)\cdot h_s(\vec{k})^* \rbrack \delta(\vec{p}+\vec{q}+\vec{k}) u_{s_p}^*(\vec{p}) u_{s_q}^*(\vec{q}),
\end{split}
\end{equation}
with $s \in \{ +,-\}$.

\begin{figure}[tbp]
\includegraphics[width=0.4\textwidth]{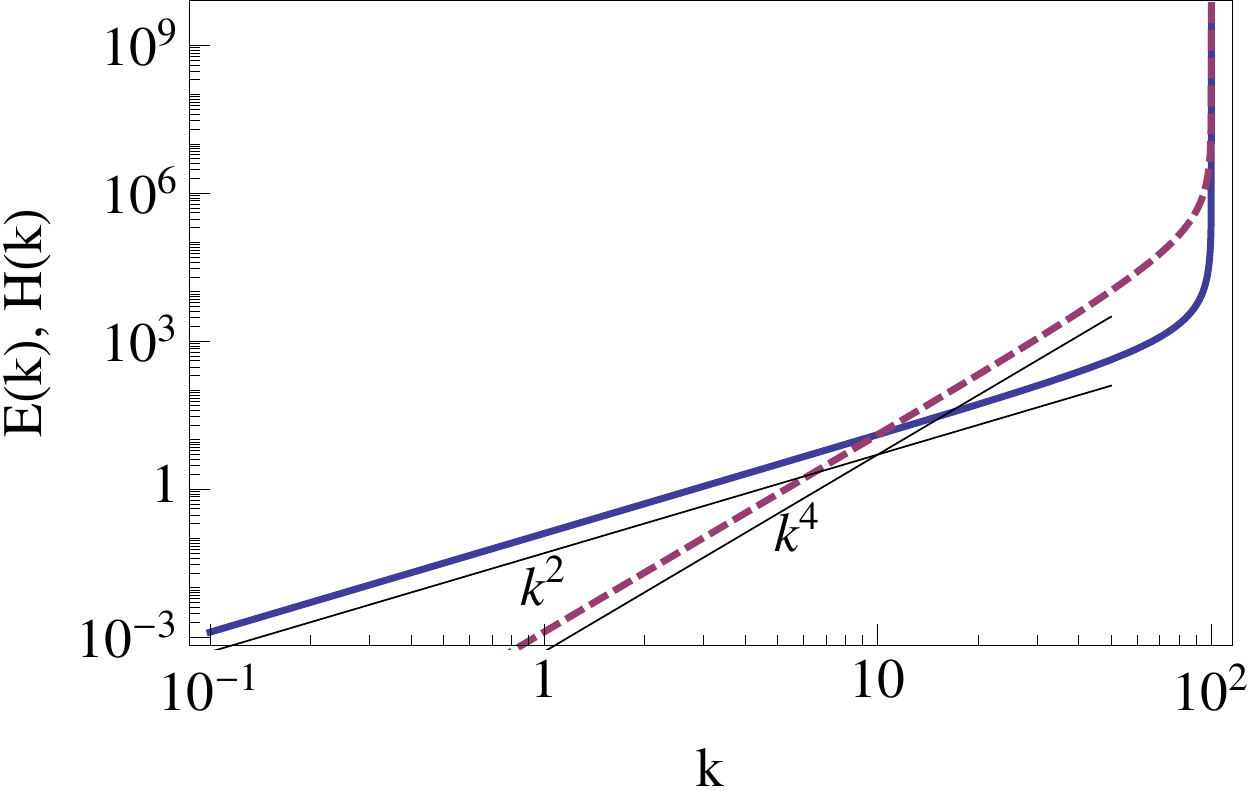}
\caption{(Color online). Energy (solid blue line) and helicity (dashed red line) spectra at statistical equilibrium for 3D turbulence. The spectra diverge for $k \to k_{max}=\beta/|\alpha|$ (here $\alpha=1$ and $\beta=100$). The thin solid lines show the $k^2$ and $k^4$ scalings corresponding to energy equipartition.}\label{3dspectrumfig}
\end{figure}

The Euler equations in 3D conserve two quadratic quantities: the energy $E$ and the helicity $H$,
\begin{align}
E &= \frac{1}{2} \int_\toruscal \vec{u}^2 = \frac{1}{2}\sum_{\vec{k} \in \pdual{\toruscal}} (\abs{u_+(\vec{k})}^2+\abs{u_-(\vec{k})}^2),\label{energydefeq}\\
H &= \frac{1}{2} \int_\toruscal \vec{u} \cdot \vort = \frac{1}{2} \sum_{\vec{k} \in \pdual{\toruscal}} k (\abs{u_+(\vec{k})}^2-\abs{u_-(\vec{k})}^2).\label{heldefeq}
\end{align}
Because they satisfy a property of \emph{detailed conservation} (they are conserved for each triadic interaction), these invariants are conserved for any spectral truncation of the system. Let us thus introduce infrared and ultraviolet cutoffs for the wave numbers: all the sums above are restricted to the set of wave numbers $\mathcal{B} = \{ \vec{k} \in \pdual{\toruscal}, k_{min} \leq k \leq k_{max}\}$. We introduce the canonical probability distribution on this truncated phase space:
\begin{equation}
\rho(\{u_+(\vec{k}),u_-(\vec{k})\}_{\vec{k} \in \mathcal{B}}) = \frac 1 {\partfun} e^{-\beta E-\alpha H},
\end{equation}
where $\beta$ and $\alpha$ are the Lagrange multipliers associated to conservation of energy and helicity, respectively, and $\partfun$ is the partition function:
\begin{align}
\partfun &= \int_{\Lambda} e^{-\beta E - \alpha H} d\mu_{\Lambda},\\
&= \prod_{\vec{k} \in \mathcal{B}} \int_0^{+\infty} da_+ \int_0^{+\infty} da_- e^{-\frac 1 2 (\beta+\alpha k)a_+^2 - \frac 1 2 (\beta-\alpha k)a_-^2},
\end{align}
where $d\mu_\Lambda$ is the Lebesgue measure on $\Lambda$.
\begin{figure*}[htbp]
\begin{tikzpicture}
\draw[-latex] (-3,0) -- (3.2,0) node[anchor=west] {$\alpha$};
\draw[-latex] (0,-1.6) -- (0,3.2) node[anchor=south] {$\beta$};
\shade[top color=blue!80!black, bottom color=white,nearly transparent] (1.5,3) -- (-1.5,3) -- (0,0) --cycle;
\draw[red!70!black,thick] (0,0) -- (-1.5,3) node[anchor=south] {$-\alpha k_{max}$};
\draw[red!70!black,thick] (1.5,3) node[anchor=south west] {$\alpha k_{max}$} -- (0,0);
\end{tikzpicture}
\hspace{2cm}
\begin{tikzpicture}
\shade[upper left=blue!80!black,lower right=blue!80!black,nearly transparent,pattern color=blue!80!black,pattern=north west lines] (3,-1.5) -- (0,0) -- (-1.5,3.) -- (3,3.) --cycle;
\shade[upper right=red!80!black, lower left=red!80!black,nearly transparent,pattern=north east lines,pattern color=red!80!black] (-3,-1.5) -- (0,0) -- (1.5,3.) -- (-3,3.) --cycle;
\draw[-latex] (-3,0) -- (3.2,0) node[anchor=west] {$\alpha$};
\draw[-latex] (0,-1.6) -- (0,3.2) node[anchor=south] {$\beta$};
\draw[blue!70!black,thick] (3,-1.5) node[anchor=east] {$-\alpha k_{min}$} -- (0,0);
\draw[blue!70!black,thick,dotted] (0,0) -- (-3,1.5);
\draw[blue!70!black,thick,dotted] (0.75,-1.5)  -- (0,0); 
\draw[blue!70!black,thick] (0,0) -- (-1.5,3) node[anchor=south] {$-\alpha k_{max}$};
\draw[red!70!black,thick,dotted] (3,1.5) -- (0,0);
\draw[red!70!black,thick] (0,0) -- (-3,-1.5) node[anchor=west] {$\alpha k_{min}$};
\draw[red!70!black,thick] (1.5,3) node[anchor=south west] {$\alpha k_{max}$} -- (0,0);
\draw[red!70!black,thick,dotted] (0,0) -- (-0.75,-1.5);
\end{tikzpicture}
\caption{
(Color online).
Accessible thermodynamic space for 3D helical flows (left: $\beta > \abs{\alpha} k_{max}$) and for phase space reduced to maximal symmetry breaking (right: $\beta \pm \alpha k_{min}>0$ and $\beta \pm \alpha k_{max}>0$ for $\Lambda_\pm$, 
marked with slanted lines and shaded in blue (resp. red) --- note that the intersection coincides with the full phase space condition $\beta > \abs{\alpha} k_{max}$). Contrary to 2D turbulence, there is no accessible negative temperature for 3D flows, but they are recovered in the restricted phase space $\Lambda_\pm$.
 }\label{canophasediagramE3Dfig}
\end{figure*}
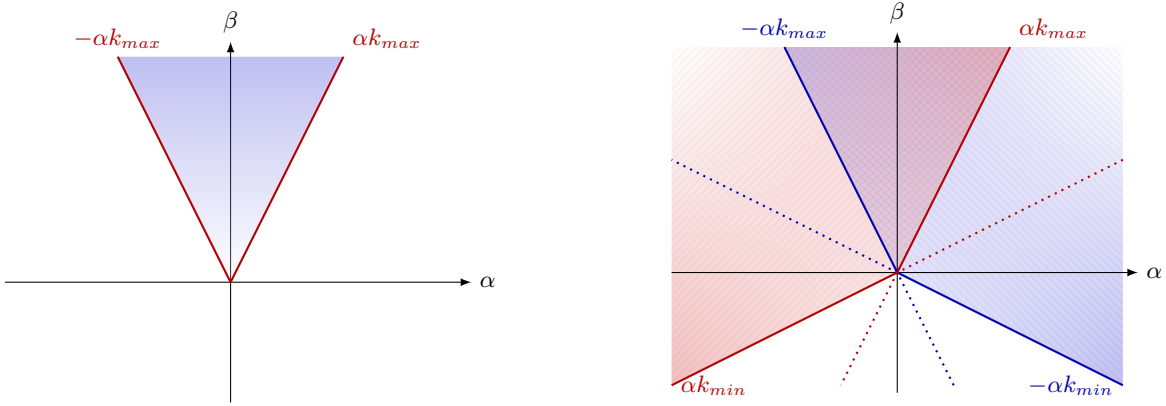
The two Gaussian integrals factor out and are easily computed; the partition function can be written as a product of two factors $\partfun=\partfun_+ \partfun_-$, with 
\begin{equation}
\partfun_\pm = \prod_{\vec{k} \in \mathcal{B}} \sqrt{\frac{\pi}{2(\beta \pm \alpha k)}}.
\end{equation}
The \emph{realizability} condition --- for the Gaussian integrals to converge --- is that $\forall \vec{k} \in \mathcal{B}, \beta \pm \alpha k >0$, which amounts to $\beta > \abs{\alpha} k_{max}$. 
In particular, the statistical temperature $\beta$ is positive, at variance with the 2D case \cite{Kraichnan1973}. The thermodynamic space of admissible values for $(\alpha,\beta)$ is represented on Fig. \ref{canophasediagramE3Dfig}.
The mean energy at statistical equilibrium is given by 
\begin{equation}
\langle E \rangle = - \frac{\partial \ln \partfun}{\partial \beta} = \sum_{\vec{k} \in \mathcal{B}} \left(\frac{1}{\beta+\alpha k} + \frac{1}{\beta-\alpha k}\right).
\end{equation}
In particular it is the sum of two contributions $\langle E_+ \rangle=\sum_{\vec{k} \in \mathcal{B}} \frac{1}{\beta+\alpha k}$ and $\langle E_- \rangle=\sum_{\vec{k} \in \mathcal{B}} \frac{1}{\beta-\alpha k}$. The resulting (isotropic) spectra of energy at statistical equilibrium, defined by 
\begin{align}
\langle E_\pm \rangle &= \int_{k_{min}}^{k_{max}} \langle E_\pm(k) \rangle dk,
\intertext{are given by}
\langle E_\pm(k) \rangle &= \frac{2\pi k^2}{\beta \pm \alpha k}, 
\intertext{so that}
\langle E(k) \rangle &= \frac{4\pi\beta k^2}{\beta^2-\alpha^2 k^2}, 
\intertext{as obtained in \cite{Kraichnan1973}. The helicity spectrum is}
\langle H(k) \rangle &=k(\langle E_+(k)\rangle - \langle E_-(k)\rangle)\\
&=\frac{4\pi \alpha k^4}{(\alpha^2 k^2-\beta^2)}. 
\end{align}
For all the admissible values of the Lagrange parameters, the energy spectrum is an increasing function of $k$, with a divergence at $k_{max}$ for $\beta=\abs{\alpha}k_{max}$ (see Fig. \ref{3dspectrumfig}).
The equilibrium energy spectrum scales as $k^2$ at low wave numbers, which correspond to the regime where energy equipartition dominates. Indeed, the energy equipartition spectrum has been observed in numerical simulations of both non-helical ($\alpha=0$, \cite{Cichowlas2005}) and helical \cite{Krstulovic2009} inviscid flows.

Assuming that there exists an energy inertial range with a constant flux of energy $\epsilon$, dimensional analysis provides the form of the energy spectrum in the inertial range: $E(k)\sim C \epsilon^{2/3} k^{-5/3}$ (Kolmogorov scaling~\cite{Kolmogorov1941a}). The slope of the energy spectrum in the inertial range is steeper than the equilibrium spectrum, and, assuming a tendency to relax towards an equilibrium state, this hints at an energy cascade towards the small scales. 
A similar reasoning could be applied to a hypothetical helicity inertial range with constant helicity flux $\eta$ (and $\epsilon=0$), but in fact the energy and helicity inertial ranges are superimposed~\cite{Brissaud1973,Andre1977}: the energy has a Kolmogorov scaling while the helicity spectrum is given by $H(k)\sim C_\eta \eta \epsilon^{-1/3}k^{-5/3}$, similarly to a passive scalar~\cite{Borue1997}. The above argument therefore remains valid, and the absolute equilibrium spectra indicates that 3D flows, even taking into account the conservation of helicity, exhibit a joint \emph{direct cascade} of energy and helicity \cite{Kraichnan1973}, in accordance with numerical simulations~\cite{Borue1997,Chen2003a}.

\section{Symmetry breaking flows and their restricted partition function}\label{symbreakflowssec}
What we have just described is the dominant behavior for homogeneous isotropic flows when taking into account the whole phase space. Now, as soon as we perturb slightly the flow and break these symmetries, different results may be obtained. For instance, we may consider breaking the parity symmetry: $P: \vec{x} \mapsto -\vec{x}$. The energy transforms as a scalar under this symmetry, while the helicity transforms as a pseudo-scalar, which implies that it vanishes for parity-invariant flows. We have seen above that imposing a sign-definite value for global helicity is not sufficient to modify the direction of the dominant energy transfers. Note that doing so does not impose a sign-definite helicity at all scales. At variance with the second quadratic invariant for 2D flows, the enstrophy $\Omega = \int \vort^2$, the helicity is not sign-definite and neither is its Fourier spectrum $H(k)$. This suggests that one should consider \emph{scale-by-scale} symmetry breaking rather than just global helicity. Indeed, in at least two different contexts, this type of constraint resulted in the emergence of an inverse cascade: numerical simulations of the Navier-Stokes equations projected on the subspace of positive definite helicity at all scales \cite{Biferale2012} and Beltrami flows (velocity-vorticity alignment) in a von Karman experiment \cite{EricHerbert2012}. These situations both correspond to cases of maximal symmetry breaking.

We will now consider submanifolds of phase space defined by such scale-by-scale helicity constraints. For each of these submanifolds, we will compute the partition function restricting the integral to the submanifold, and study the resulting equilibrium statistical properties.

\subsection{Maximal symmetry breaking}\label{maxsymbreaksec}
\begin{figure*}[tbp]
\includegraphics[width=\textwidth]{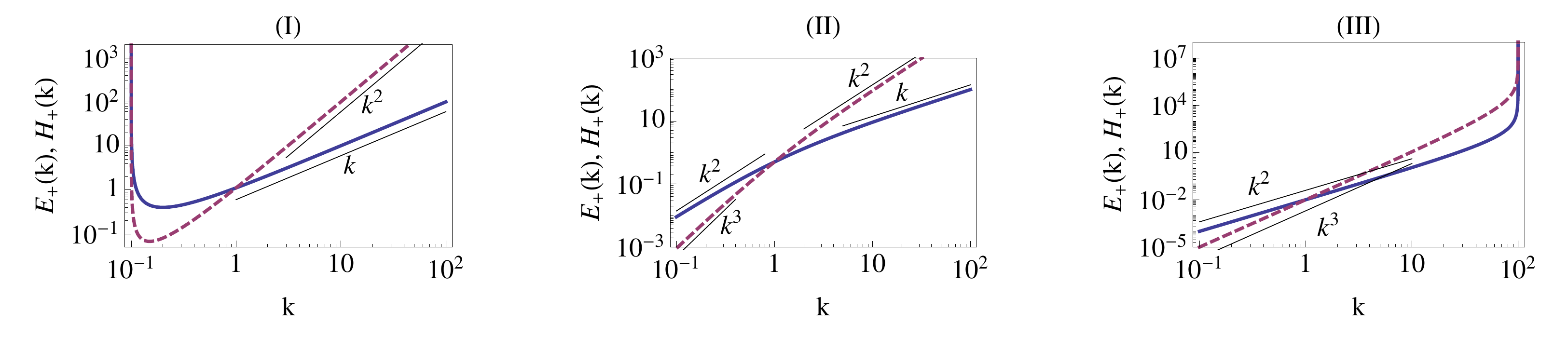}
\caption{(Color online). Equilibrium energy (solid blue lines) and helicity (dashed red lines) spectra $(E_+(k), H_+(k))$ for the different $(\alpha,\beta)$ regimes, for restricted phase space $\Lambda_+$. 
Left: $\alpha>0, \beta <0$ (regime I); the spectra have a well shape, with an infrared divergence and an increase at large $k$ (the thin lines indicate the $k$ and $k^2$ scaling). 
Middle: $\alpha>0, \beta>0$ (regime II); the spectra increase with $k$, with scalings $(k^2,k^3)$ at low-$k$ and $(k,k^2)$ at large $k$. Right: $\alpha<0,\beta>0$ (regime III); the spectra increase as $k$ increases, and there is an ultraviolet divergence (the thin lines indicate the $(k^2,k^3)$ scalings at low $k$).}\label{3ddecspectrafig}
\end{figure*}
The Cauchy-Schwartz inequality yields the constraint $\abs{H(k)} \leq k E(k)$. The equality corresponds to maximum local symmetry breaking. From Eqs. \ref{energydefeq,heldefeq}, we see that it occurs on the submanifold in phase space defined by $a_+(\vec{k})=0$ (for $H(k)=-kE(k)$) or $a_-(\vec{k})=0$ (for $H(k)=kE(k)$).
Hence, it is natural to consider first the restricted partition function corresponding to the submanifolds of phase space 
\begin{align}
\Lambda_+ &= \{ u \in \Lambda, \forall \vec{k} \in \mathcal{B}, u_{-}(\vec{k})=0\} \subset \Lambda,\\
\Lambda_- &= \{ u \in \Lambda, \forall \vec{k} \in \mathcal{B}, u_{+}(\vec{k})=0\} \subset \Lambda.
\end{align}
Note that $\Lambda=\Lambda_+ \oplus \Lambda_-$. These restricted partition functions are 
\begin{align}
\partfun_\pm &=\int_{\Lambda_\pm} e^{-\beta E - \alpha H} d\mu_{\Lambda_\pm},\\
&=\prod_{\vec{k} \in \mathcal{B}} \sqrt{\frac{\pi}{2(\beta \pm \alpha k)}},
\end{align}
as already computed above. The \emph{realizability} condition reduces to $\beta\pm \alpha k_{min} >0, \beta \pm \alpha k_{max} >0$ (see Fig. \ref{canophasediagramE3Dfig}).	 This condition is reminiscent of that of 2D turbulence \cite{Kraichnan1967}, $\beta+\alpha k_{min}^2>0, \beta+\alpha k_{max}^2>0$, where $\alpha$ is the Lagrange parameter associated to conservation of enstrophy in that case. Like in 2D flows~\cite{Kraichnan1975,Kraichnan1980}, we now recover \emph{negative temperature states} for each of the restricted phase spaces $\Lambda_\pm$ and we can identify three regimes --- for $\Lambda_+$:
\begin{enumerate}[label=(\Roman{*})]
\item $\beta<0, \alpha>0$,
\item $\beta>0, \alpha>0$,
\item $\beta>0,\alpha<0$.
\end{enumerate}
The $\Lambda_-$ case follows from changing the sign of $\alpha$.

These regimes can also be characterized in terms of a characteristic length scale $k_c=H/E$: 
\begin{enumerate}[label=(\Roman{*})]
\item $k_{min} \leq k_c \leq k_a$,
\item $k_a \leq k_c \leq k_b$,
\item $k_b \leq k_c \leq k_{max}$,
\end{enumerate}
with $k_a=2/3 \times (k_{max}^3-k_{min}^3)/(k_{max}^2-k_{min}^2)$ and $k_b=3/4 \times (k_{max}^4-k_{min}^4)/(k_{max}^3-k_{min}^3)$. We show in Fig. \ref{3ddecspectrafig} the typical equilibrium spectra for energy and helicity corresponding to these three cases. Regime (III) is a high-helicity regime with concentration of the energy at the small scales, similar to the full phase space of 3D helical turbulence (compare to Fig. \ref{3dspectrumfig}), apart from the $k^3$ helicity scaling instead of $k^4$. In the intermediate regime (II), we have two ranges corresponding to energy equipartition at large scales and helicity equipartition at small scales. In addition to helicity equipartition at small scales, regime (I) features an energy \emph{condensation at large scales} as in the 2D case.
 This energy condensation at statistical equilibrium for the ideal system indicates that if there is an energy inertial range in the forced dissipative system, it should correspond to an inverse cascade. Indeed, in an inertial range with constant energy flux $\epsilon$ and zero helicity flux $\eta=0$, dimensional analysis gives an energy spectrum scaling as $E(k) \sim C_\epsilon \epsilon^{2/3} k^{-5/3}$ (as in 2D turbulence). In particular, the energy spectrum in the inertial range should be shallower than the equilibrium. Hence, assuming a tendency to relax towards the equilibrium, even in the presence of forcing and dissipation, the direction of the cascade should be towards the larger scales.
The existence of the energy inertial range and the energy cascade towards the large scales have been observed in both numerical simulations retaining only same helical polarization interactions \cite{Biferale2012}, which amounts to projecting the dynamics on $\Lambda_\pm$, and a von K\'arman experiment~\cite{EricHerbert2012} where the system reaches a steady state which is a Beltrami flow~\cite{Naso2010b}.
In a similar manner, the equilibrium spectrum in regime (I) points to a tendency towards helicity equipartition at small scales, which implies a positive flux of helicity.
Therefore, the statistical mechanics approach seems consistent with the general idea that a positive-definite helicity plays an analogous role to enstrophy in 2D flows~\cite{Kraichnan1973,Biferale2012}, in that it prevents a simultaneous downscale cascade of energy and helicity.

 \begin{figure*}[tbp]
\includegraphics[width=\textwidth]{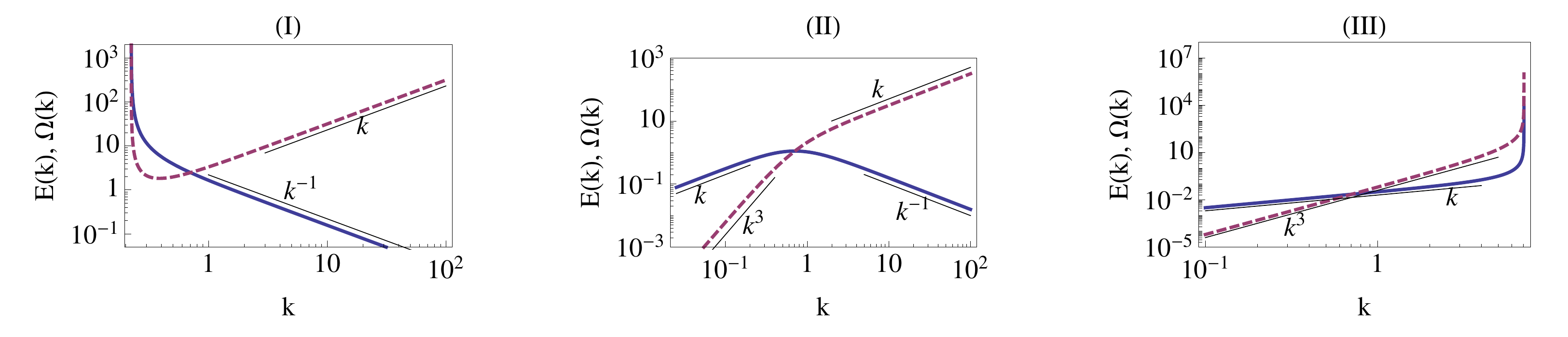}
\caption{(Color online). Equilibrium energy (solid blue lines) and enstrophy (dashed red lines) spectra $(E(k), \Omega(k))$ for the different $(\alpha,\beta)$ regimes in 2D Turbulence. 
Left: $\alpha>0, \beta <0$ (regime I); the spectra have an infrared divergence (the thin lines indicate the $k$ and $k^{-1}$ scaling). 
Middle: $\alpha>0, \beta>0$ (regime II); the energy spectrum has a bell shape, the enstrophy spectrum increases with $k$, with scalings $(k,k^3)$ at low-$k$ and $(k^{-1},k)$ at large $k$. Right: $\alpha<0,\beta>0$ (regime III); the spectra increase as $k$ increases, and there is an ultraviolet divergence (the thin lines indicate the $(k,k^3)$ scalings at low $k$).}\label{2dspectrafig}
\end{figure*}
To describe further the analogy, we show in Fig. \ref{2dspectrafig} the equilibrium energy and enstrophy spectra obtained in the three regimes analogous to (I), (II) and (III) in 2D turbulence~\cite{Kraichnan1980}. The three regimes look quite similar, with a high enstrophy regime where the energy concentrates at the small scales (III), an intermediate regime (II), and a low enstrophy regime with energy condensation at the large scales (I). 
Nevertheless, there are also interesting differences. 
First, in the energy equipartition spectra, the enstrophy (2D) and the helicity (3D) spectra have the same scalings, but of course the energy spectrum does not have the same scaling in the 2D and the 3D case. This is simply due to the fact that the number of modes in a spherical shell of radius $k$ scales as $k^{D-1}$ where $D$ is the dimension. But this does not change the qualitative behavior; in particular, these spectra remain increasing functions of $k$.
More interestingly, the helicity equipartition regime on $\Lambda_+$ and the enstrophy equipartition regime in 2D do not yield the same scalings for the energy and helicity/enstrophy spectra. In particular, the energy spectrum at enstrophy equipartition in 2D decreases as $k^{-1}$, while at helicity equipartition on $\Lambda_+$, it increases as $k$. In other words, the energy spectrum at helicity equipartition on $\Lambda_+$ behaves like the enstrophy spectrum at enstrophy equipartition in 2D.
This indicates that in spite of the presence of an inverse cascade, substantial transfers of energy to the small scales should nevertheless be expected. 

This is at variance with the 2D case, for which such forward energy transfers are constrained to be very small by the Fj{\o}rtoft argument \cite{Fjortoft1953} (see also Refs. \onlinecite{Rhines1979,NazarenkoBook}). In its \emph{centroid} version, the Fj{\o}rtoft argument relies on simple inequalities between the energy containing wavenumber $k_E=\int_0^{+\infty} k E(k)dk/E$ and the enstrophy containing wavenumber $k_\Omega=\int_0^{+\infty} k \Omega(k)dk/\Omega$. The Cauchy-Schwarz inequality implies that $k_E \leq \sqrt{\Omega/E}, k_\Omega \geq \sqrt{\Omega/E}$ and $k_E k_\Omega \geq \Omega/E$. Therefore, the energy containing wavenumber is bounded from above, the enstrophy containing wavenumber is bounded from below, and their product is bounded from below. Hence, a decrease of $k_E$ (i.e. inverse cascade of energy) must be accompanied by an increase of $k_\Omega$ (i.e. direct cascade of enstrophy). Introducing the $\ell$-centroids $\ell_E=\int_0^{+\infty} k^{-1} E(k)dk/E, \ell_\Omega=\int_0^{+\infty} k^{-1} \Omega(k)dk/\Omega$ and deriving the analogous inequalities $\ell_E\geq \sqrt{E/\Omega}, \ell_\Omega \leq \sqrt{E/\Omega}$ and $\ell_E \ell_\Omega \geq E/\Omega$, we obtain the converse of this statement: the direct cascade of enstrophy implies an inverse cascade of energy.

Here, the relation between the two invariants does not yield such a strong constraint; with the same definition for the energy containing scale $k_E$, we see that $k_E=H/E=k_c$ is constant in time. Similarly, introducing the characteristic wave numbers and length scales $k_H, \ell_E$ and $\ell_H$, we see that $\ell_H=k_E^{-1}$ is also constant, and that $k_H \geq k_E$ and $\ell_E \geq \ell_H$. The inequalities indicate that there is a lower bound to the helicity containing wave number and  to the energy containing scales, analogously to 2D turbulence. But the fact that $k_E$ is constant in time points out that an increase of $E(k)$ for $k$ smaller than $k_E$ (i.e. at large scales) must be compensated by an increase for $k$ larger than $k_E$ (i.e. at small scales), so that the mean value of the distribution remains constant in time. A similar reasoning holds with $H(k)$, as $\ell_H$ is also time independent. 

It remains unclear if and how this property of the ideal system will impact the forced-dissipative case. It may indicate that in addition to the energy inverse cascade, substantial forward energy transfers may remain, although it seems unlikely that they should organize into an inertial range with a constant flux. Note that the possibility of a \emph{dual cascade} phenomenology (coexisting inverse and direct cascades for the same quantity) has been suggested in rotating~\cite{Mininni2010a} and rotating-stratified turbulence~\cite{PouquetArxiv2013}, but in the case of helically decimated flows, the possibility of a range of constant positive energy fluxes has been ruled out by Biferale et al.~\cite{Biferale2012,Biferale2013b}.

\subsection{Infinitesimal Symmetry breaking}\label{infsymbreaksec}
In the previous paragraph, we have shown that an inverse cascade of energy could be obtained as a subdominant contribution to the partition function stemming from the part of phase space corresponding to maximal parity symmetry breaking: on $\Lambda_\pm$, the absolute value of the relative helicity $\sigma(k)=H(k)/kE(k)$ is 
equal to its maximum value of 1 at all scales. A legitimate question to ask is: how much symmetry breaking do we need to obtain this phenomenon?
Let us consider a different family of restricted phase spaces: 
\begin{align}
\Lambda_\varepsilon = \{ u \in \Lambda, \forall \vec{k} \in \mathcal{B}, u_+(\vec{k})=\sqrt{1+\varepsilon}u_-(\vec{k})\} \subset \Lambda. 
\end{align}
These flows have a relative helicity $\sigma(k)=\frac{\varepsilon}{2+\varepsilon}$, which goes to zero uniformly as $\varepsilon$ does. The restricted partition function can easily be computed: 
\begin{align}
\partfun_\varepsilon &= \int_{\Lambda_\varepsilon} e^{-\beta E - \alpha H} d\mu_{\Lambda_\varepsilon},\\
&=\prod_{\vec{k} \in \mathcal{B}} \sqrt{\frac{\pi}{2(\beta(2+\varepsilon)+\varepsilon \alpha k)}}.
\end{align}
The \emph{realizability} condition is $\beta>-\varepsilon \alpha k_{min}/(2+\varepsilon)$ and $\beta>-\varepsilon \alpha k_{max}/(2+\varepsilon)$, which yields a thermodynamic space analogous to Fig. \ref{canophasediagramE3Dfig} (right), and the three regimes identified in the previous paragraph remain valid here, with the same associated spectra (see Fig. \ref{3ddecspectrafig}). In particular, the inverse cascade regime still exists in spite of an infinitesimal symmetry breaking. Nevertheless, the extent of the region of thermodynamic space corresponding to this regime also goes to zero as the amount of symmetry breaking ($\varepsilon$, or equivalently, $\sigma(k)$) does.
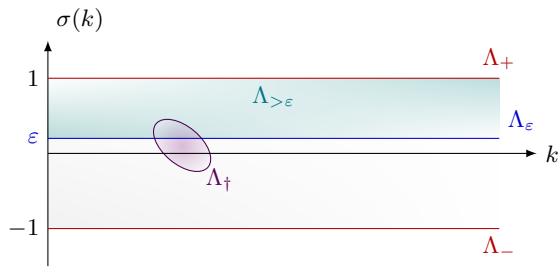
\begin{figure}[tbp]
\begin{tikzpicture}
\shade[lower left=gray!10, upper right=white] (-3,-1) -- (3,-1) -- (3,1) -- (-3,1);
\shade[lower left=blue!50!green, upper right=blue!50!green,nearly transparent] (-3,0.2) -- (3,0.2) -- (3,1) -- (-3,1) node[blue!55!green,opaque,midway,below] {$\Lambda_{>\varepsilon}$};
\shade[inner color=violet!90!black, outer color=violet!10!white,rotate=-40,nearly transparent,draw=violet!60!black,draw opacity=1] (-1,-0.7) ellipse [x radius=0.45cm, y radius=0.25cm] node[below right=2mm,opaque,violet!70!black] {$\Lambda_\dagger$};
\draw[-latex] (-3,0) -- (3.5,0) node[anchor=west] {$k$};
\draw[-latex] (-3,-1.5) -- (-3,1.5) node[anchor=south west] {$\sigma(k)$};
\draw[red!70!black] (-3,1) node[black,anchor=east] {$1$} -- (3,1) node[anchor=south] {$\Lambda_+$};
\draw[red!70!black] (-3,-1) node[black,anchor=east] {$-1$} -- (3,-1) node[anchor=north] {$\Lambda_-$};
\draw[blue!80!black] (-3,0.2) node[anchor=east] {$\varepsilon$} -- (3,0.2) node[anchor=south west] {$\Lambda_\varepsilon$};
\end{tikzpicture}
\caption{(Color online). Scale-by-scale relative helicity for the different restricted phase spaces considered. For the total phase pace $\Lambda$, 
$|\sigma(k)|\le 1$,
while $\Lambda_\pm$ denote the part of phase space where 
$\sigma(k)=\pm1$;
$\Lambda_\varepsilon$ corresponds to a constant infinitesimal value $\varepsilon$ and $\Lambda_{>\varepsilon}$ is the largest restricted phase space for which 
$\sigma(k)>0$.
As soon as the restricted phase space intersects both regions of positive and negative relative helicity (e.g. $\Lambda_\dagger$ on the graph), the inverse cascade breaks down.
}\label{relhelfig}
\end{figure}
More generally, we can define the restricted phase space 
\begin{align}
\Lambda_{>\varepsilon} = \{ u \in \Lambda, \forall \vec{k} \in \mathcal{B}, \abs{u_+(\vec{k})}>\sqrt{1+\varepsilon}\abs{u_-(\vec{k})}\} \subset \Lambda,
\end{align}
or analogously, $\Lambda_{<\varepsilon}$. For any subset of this restricted phase space, the helicity is positive definite. Besides, the realizability condition for the restricted partition function will always be satisfied by $\alpha$ and $\beta$ such that $\beta+\alpha k_{min} >0$ and $\beta+\alpha k_{max}>0$, which means that the negative temperature states are attainable, and the inverse cascade regime (I) is possible. On the contrary, as soon as a subset of phase space $\Lambda_\dagger$ has a closure which intersects two of the three subsets $\Lambda_0,\Lambda_{>0},\Lambda_{<0}$ (note that $\Lambda=\Lambda_0\cup\Lambda_{>0}\cup\Lambda_{<0}$), the realizability condition implies $\beta>0$, and the inverse cascade regime vanishes. This shows that the only restricted phase spaces leading to an inverse cascade are those for which the relative helicity is sign-definite at all scales (see Fig. \ref{relhelfig}).

\section{Conclusion}

In this paper, we have used a tool from statistical mechanics, the \emph{restricted partition function}, to finely probe the most probable outcome of the nonlinear interactions in turbulent flows, especially in terms of energy transfer and energy cascade. 
This method can be useful in all cases where taking into account the whole phase space is not relevant, for instance because the dominant modes in the ideal system will in fact not be dominant in the real system where the particular choice of forcing will put more emphasis on a subset of phase space, even if the restricted phase space is not rigorously an invariant manifold for the dynamics. As an example, we have shown using this method that perturbing slightly 3D homogeneous isotropic turbulence by imposing a symmetry breaking --- here breaking the parity symmetry, measured by helicity --- yields richer phenomenologies than the standard 
cascade. This result provides analytical support to recent numerical simulations \cite{Biferale2012} and experiments \cite{EricHerbert2012} for Beltrami flows. 
We have further shown that an \emph{infinitesimal} symmetry breaking at all scales is sufficient to obtain the inverse energy cascade, and that this behavior is characteristic of restricted phase space for which the relative helicity is sign-definite at all scales. A sharp transition to a regime of direct energy cascade is expected when the sign constraint on relative helicity is released, even in the slightest manner.
Therefore, the results of this statistical mechanics approach restricted to submanifolds of phase space tends to confirm the idea that when helicity becomes sign-definite at all scales, it plays an analogous role to enstrophy in 2D and forbids the simultaneous existence of a forward cascade of energy and helicity, as anticipated by Kraichnan~\cite{Kraichnan1973} and further developed by Biferale et al.~\cite{Biferale2012,Biferale2013b}. We have further discussed this analogy with 2D turbulence by examining the differences between the two systems, in the helicity/enstrophy equipartition spectra on the one hand, and in the constraints enforced by the two invariants through the Fj{\o}rtoft argument on the other hand.

The methods exposed here are likely to bear fruits in other systems where the symmetry is broken as a result of a physical force, like rotation or stratification~\cite{Herbert2013e}, which prevail in geophysical flows and break the continuous part of the symmetry group. Indeed, in rotating and/or stratified turbulence, it is customary to introduce different decompositions of phase space in terms of \enquote{2D modes}, which are expected to dominate at large (horizontal) scales, while \enquote{3D modes} prevail at scales smaller than the Zeman or Ozmidov scale at which isotropy is recovered \cite{Mininni2012}. Although the large-scale, balanced modes of geophysical flows do not form rigorously an invariant manifold of the full phase space, in practice the forcing does not excite all modes equally, and the dynamics may remain in the vicinity of this \emph{slow manifold} \cite{Lorenz1986}. 
Restricted partition functions provide a way to study analytically the relative roles of these subsets of phase space.

\begin{acknowledgments}
The National Center for Atmospheric Research is sponsored by the National Science Foundation. Numerous and insightful discussions with Annick Pouquet are warmly acknowledged.
\end{acknowledgments}

\end{document}